\documentclass[final,3p,times]{elsarticle}

\usepackage[utf8]{inputenc}
\usepackage{graphicx}
\usepackage{dcolumn}
\usepackage{bm}
\usepackage{amssymb}
\usepackage{epstopdf}
\usepackage{amsmath}
\usepackage{xcolor}
\usepackage{enumerate}
\usepackage{lineno,hyperref}
\usepackage{setspace}
\usepackage{natbib}

\journal{Ecological Modelling (09 November 2018)}

\begin{document}

\begin{frontmatter}

   %
\title{Early warning signals in plant disease outbreaks}
\author{S.~Orozco-Fuentes$^\text{a}$, G. Griffiths$^\text{a}$, M. J. Holmes$^\text{b}$, R. Ettelaie$^\text{b}$, J. Smith$^\text{b}$, A. W. Baggaley$^\text{a}$ and N. G. Parker$^\text{a}$}

\address{$^\text{a}$School of Mathematics, Statistics and Physics, Newcastle University, Newcastle upon Tyne, NE1 7RU, UK}
\address{$^\text{b}$School of Food Science and Nutrition, University of Leeds, Leeds, LS2 9JT, UK}




\date{DOI: 0.1016/j.ecolmodel.2018.11.003}

\begin{abstract}

Infectious disease outbreaks in plants threaten ecosystems, agricultural crops and food trade. Currently, several fungal diseases are affecting forests worldwide, posing a major risk to tree species, habitats and consequently ecosystem decay. Prediction and control of disease spread are difficult, mainly due to the complexity of the interaction between individual components involved. In this work, we introduce a lattice-based epidemic model coupled with a stochastic process that mimics, in a very simplified way, the interaction between the hosts and pathogen. We studied the disease spread by measuring the propagation velocity of the pathogen on the susceptible hosts. Our quantitative results indicate the occurrence of a critical transition between two stable phases: local confinement and an extended epiphytotic outbreak that depends on the density of the susceptible individuals.
Quantitative predictions of epiphytotics are performed using the framework early-warning indicators for impending regime shifts, widely applied on dynamical systems.  These signals forecast successfully the outcome of the critical shift between the two stable phases before the system enters the epiphytotic regime. Our study demonstrates that early-warning indicators could be useful for the prediction of forest disease epidemics through mathematical and computational models suited to more specific pathogen-host-environmental interactions. Our results may also be useful to identify a suitable planting density to slow down disease spread and in the future, design highly resilient forests.

\end{abstract}

\end{frontmatter}




\section{Introduction}\label{sec: Intro}

Invasive non-indigenous pathogens and vectors such as fungi, bacteria and insects pose a serious threat to trees and forest health worldwide. The recent and well-publicised outbreak of the ash dieback fungus ({\it Hymenoscyphus pseudoalbidus}) and emerald ash borer ({\it Agrilus planipennis}) risks the survival of the ash tree ({\it Fraxinus excelsior}) in the UK, one of the most abundant trees in small woodlands and high forests across the country \citep{AshTreeDistro,forestryNFIEng}. At the same time, this fungus, threatens the ash tree extinction across the European continent \citep{Mitchell201495, JEC:JEC12566,MPP:MPP12073}. The larch tree disease, caused by the fungus {\it Phytophthora ramorum} ({\it P. ramorum}), continues to spread through conifer forests in both Scotland and Wales, changing the landscape and forcing the Forestry Commission to fell thousands of hectares of trees to slow down the spread of the disease \citep{forestryPRamorum}.

Historically, these events cause catastrophic ecological, economic and social impact, and motivate a detailed understanding of the mechanisms that underlie the epidemics, from which strategies to manage and prevent future occurrences can be developed systematically \citep{HarwoodPotter2010}. The propagation of these infectious agents to the susceptible trees depends on a plethora of biological, geographical, climatic and anthropological factors. In the literature, several spatial models have been developed for forests diseases, that consider specific factors aiding the dispersal of invasive pests, such as vectors (insects, humans) or economical activities like the international plant trade and timber industry \citep{MACNADBAY2004249, Alfinitosrep27202, HarwoodPotter2010}. However, these models, specifically designed to account for large geographical areas of natural forests, are very complex and require a large amount of input data to predict the disease spread.

\begin{figure*}
  \centering
    \includegraphics[width=0.95\textwidth]{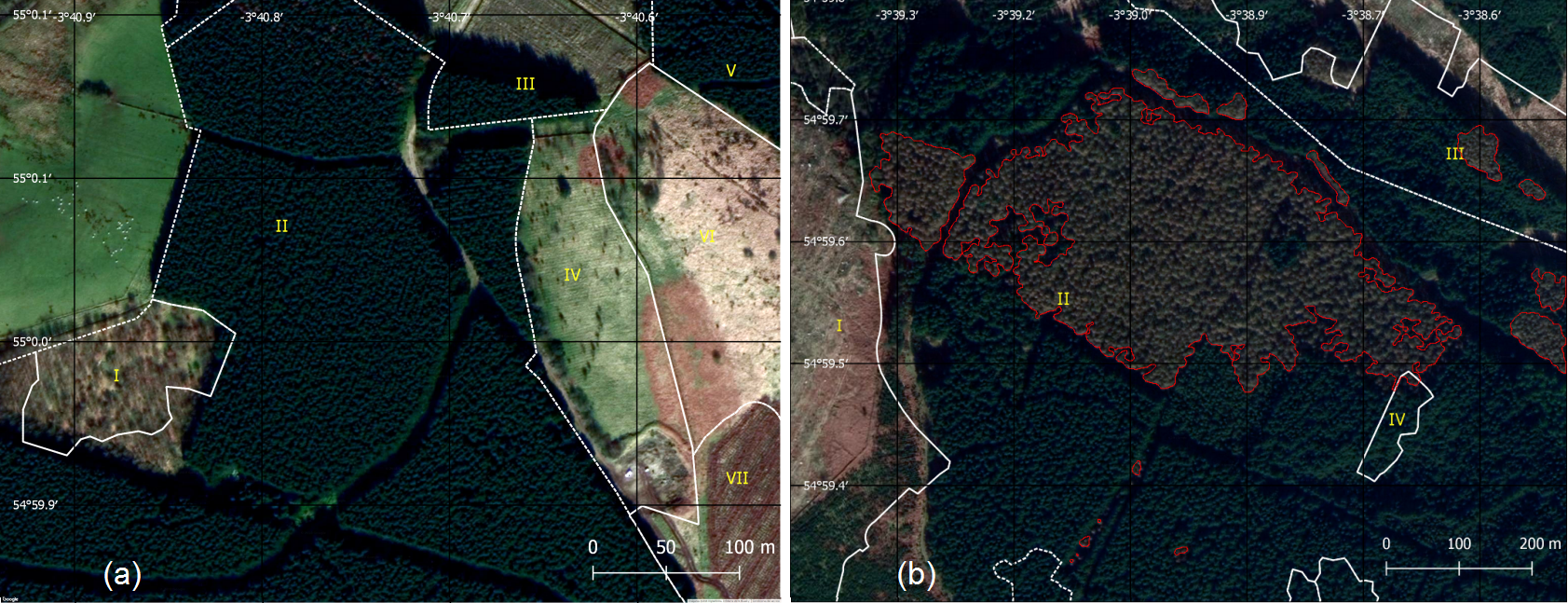}
    \caption{Woodland patches located in the Mabie Forest near Dumfries and Galloway, UK. (a) Monocultures of coniferous forests outlined according to the interpreted forest type (IFT) accounted in the National Forest Inventory Scotland 2016 (white dotted lines): I, young trees; II and V conifer trees; III and VII, felled area; IV, bare area and VI, grassland. Trees were planted at intervals of 2-3 m. (b) Coniferous forests with patches of trees infected with {\it P. ramorum} (red outline). The IFTs showed are I, felled trees; II, conifer trees, III, young trees and IV, broadleaved trees. Maps are showed in latitude/longitude coordinates and  were obtained with QGIS 2.18 'Las Palmas', using \copyright2018 Google Satellite datasets. To account for the interpreted forest types we used the National Forest Inventories from Scotland (2016) \citep{forestryNFIScotland}. Image analysis was done with ImageJ \citep{Imagej}.}
    \label{forestsUK}
\end{figure*}

The forests in the United Kingdom and several parts of Europe have been reshaped continually since the mid-Holocene due to anthropogenic factors \citep{forestIndustrial}. During the last centuries, the timber industry has left a characteristic homogeneous pattern in the woodland patches: forests managed for timber are usually planted in lines or curved lines. These patterns allow an efficient management and an even access to sunlight and nutrients; with all trees in the plantations being even-aged monocultures of conifer or broadleaved forests \citep{forestryShaped}, see figure \ref{forestsUK}(left). However, the homogeneity in the trees diminishes the resilience of the forests to several threats, including forest diseases \citep{rist-moen}. Examples of this can be observed in the current outbreak of {\it P. ramorum} spreading in the Mabie Forest in Scotland, see figure \ref{forestsUK}(right), which consists mainly of coniferous forests with trees planted at the same time between 2-3 m of each other.

Mathematical modelling provides a powerful approach to understand, predict and counter-act disease propagation \citep{bate-white, macpherson-hanley} with the advantage of fine-tuning the model to take into account specific attributes found in the forests. In this work we have developed an individual-based model in which trees are represented explicitly alongside their susceptibility to disease and infectious status, to account for disease spread in terms of the tree density, a basic forest measurement calculated in the field. Our model is similar in nature to the forest fires and percolation lattice models widely investigated in the literature in which transmission occurs upon direct contact \citep{PerBak,BeerEnting,Grassberger93} and there is only spatial stochasticity. However, to consider the effect of a simultaneous presence of both spatial and temporal stochasticity we introduced a probability of transmission for the trees to catch the disease, which is not considered in the former models.

Early-warning indicators for abrupt changes in the behaviour of complex systems, group a set of statistical properties measured on parameters that change in unique ways before the occurrence of a catastrophic shift, (also known in the literature as tipping point or critical transition), which occurs when a system switches abruptly between alternate equilibria \citep{schefferNature,scheffer-book}. These indicators are generic and suitable for application across many system types, even when the underlying system dynamics are poorly understood \citep{Brock2006, schefferNature,moralesfrank}.

In ecosystems, early-warning methods have been used to predict the occurrence of desertification processes \citep{Corrado2014}, animal extinction in deteriorating environments \citep{Griffen2010}, behaviour of aquatic ecosystems \citep{Gsell13122016}, and have been applied on climate models for the simulation of dieback on the Amazon rainforest \citep{Boulton2013}. Most recently, their applicability as an effective model to monitor tree mortality has been highlighted in \citep{Rogers2018} through satellite data. Therefore, their usefulness to predict changes on degradation processes for biological systems has increased in the literature during the last decade \citep{Brock2006,schefferNature,dakos2012,Hunsicker2016,Ravi2017}. The main purpose behind these indicators is their effectiveness to identify properties in an ecological system that would change significantly as it approaches a tipping point between different stable states. However, this idea has been applied only to a handful of ecological problems due to the unavailability of data sets, \citep{Gsell13122016, Hunsicker2016, Ravi2017}.

Following this premise, our interest lies in analysing the dynamics of disease outbreaks under the scope of classical early-warning techniques using the tree density as a state variable. This suggests applying the universality class and scaling exponents---which have been widely studied in the literature \citep{bunde-havlin,stauffer}---to the prediction and detection of the transition from the progression of the disease to an epiphytotic outbreak.



The structure of the paper is as follows. In \S\ref{sec:model}, we propose a simplified model of a forest, in terms of susceptible, infected or removed individuals (SIR model) which exhibits a phase transition above a percolation threshold \citep{Grassberger93,ClarSchwabl,bunde-havlin}. In \S\ref{sec:results}, we show the results of the simulations in which, we quantify the propagation velocity of the infection. Then, we obtain the phase diagram of the contained-to-outbreak phases of the system and study the behaviour of relevant parameters. Finally, in \S\ref{sec:discussion} we discuss the meaning and implications of our findings.

\section{Materials and methods}\label{sec:model}

\begin{figure}
  \centering
    \includegraphics[width=0.46\textwidth]{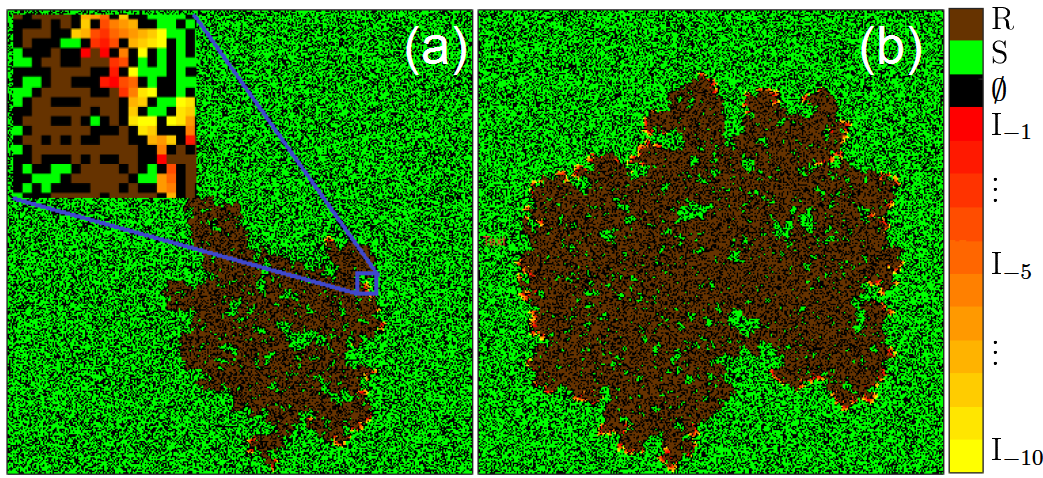}
    \includegraphics[width=0.41\textwidth]{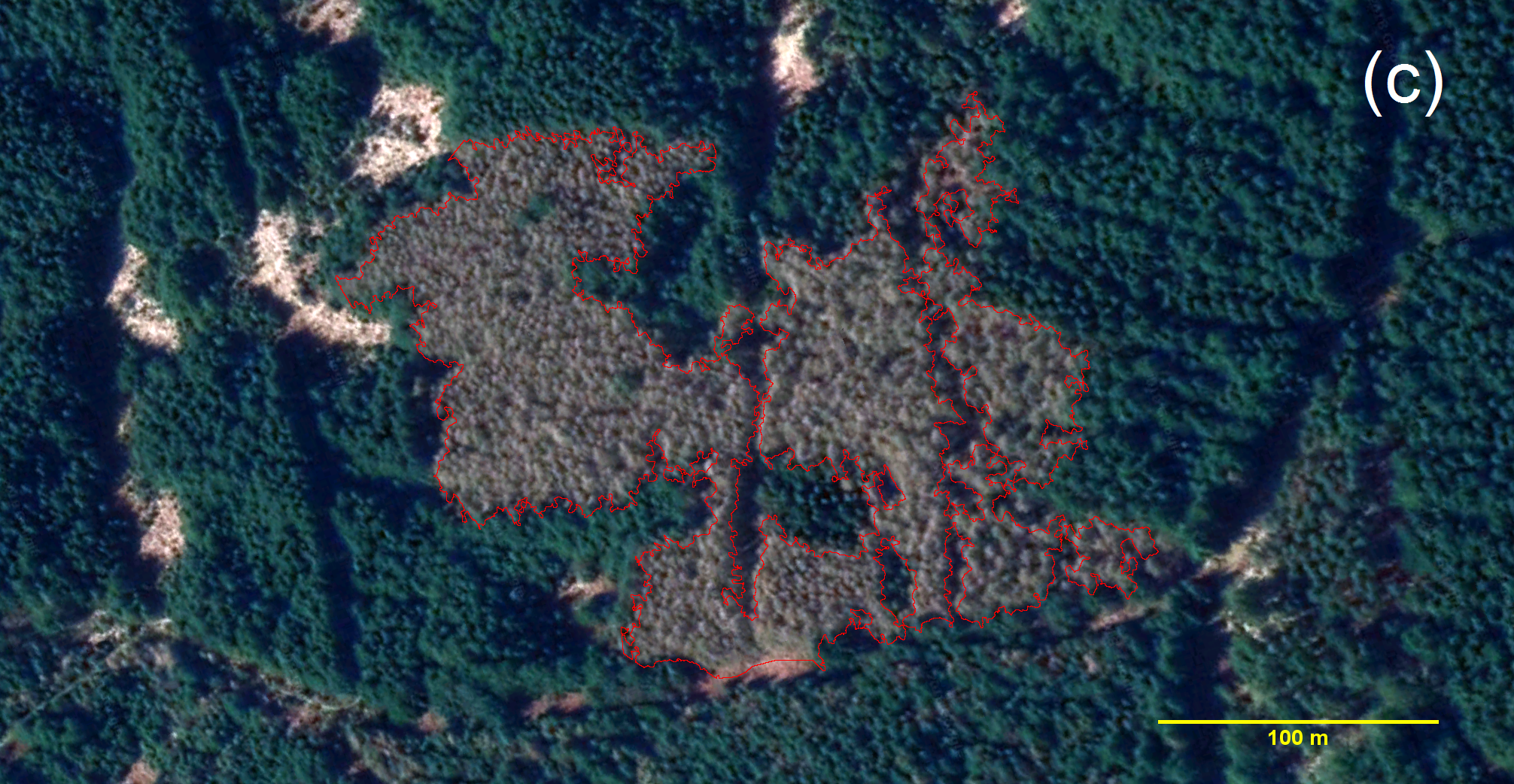}
    \caption{Typical final configurations of disease spread obtained for a system with $L = 500$, $\beta=0.5$ and tree densities $\rho=$ 0.60 (a) and 0.62 (b). Near the critical transition, slight changes in tree densities result in different spatial patterns of disease spread. {\it Inset:} Detail of the sites statuses with trees represented following the colour bar showed on the right: empty ($\emptyset$), susceptible (S), removed (R) and infected (I). The simulations were stopped after the disease dies out (a) or when it reached the edge of the system (b). (c) Patterns of disease spread ({\it P. ramorum}) on coniferous trees in the Mabie Forest near Dumfries, Scotland, UK. The total forest patch area is $\sim$ 250 ha and the infected region $\sim$ 2.5 ha. \copyright2018 Google Satellite datasets. Image analysis was done with ImageJ \citep{Imagej}.}
    \label{L500}
\end{figure}

We model a forest as a regular square lattice of dimensions $L \times L$ where $L=500$, see Fig.~\ref{L500}(a-b).  The forest landscape is flat and there is only one type of vegetation, with the initial occupation of trees following a Bernoulli trial according to a binomial distribution with mean $\rho L^2$, in which each trial has two possible outcomes a tree or an empty space. Forest patches with monocultures of trees with the same age that fullfill these characteristics can be found in several regions in Scotland, UK, see figure \ref{L500}(c), in which the trees planted every 2-3 m are infected with {\it P. ramorum}.

Following this, in our model, each site can exist in one of four states: susceptible ($S$), infected ($I$), removed ($R$) and empty ($\emptyset$). Susceptible individuals correspond to a single or several trees distributed randomly which can become infected. An infected site, represents a patch of vegetation that has acquired a terminal disease, a removed site corresponds to the space left by the infected site after the vegetation dies, and empty sites correspond to regions in the landscape where no susceptible vegetation can grow, see figure \ref{L500}(a-b). The location of the forest patches is constant in time, such that vegetation sites (either $S$, $I$ or $R$) are randomly distributed with a density $\rho$.

Once a site in the $S$ category acquires an infection, its status is changed to $I$ and a numerical label $\eta$ is attached to it. $\eta$ increases with time at a constant pace, ranging from $-T$ to $0$. Whereupon at $\eta=0$, the tree at the site dies and is removed, i.e., its status is changed to $R$. Henceforth, $T$ corresponds to the infectious period, that is, the time in which an infected tree can transmit the infection.

After a tree is infected, it has a probability $\beta$ of transmitting the disease to a neighbouring susceptible site during the infectious period. In epidemiological terms, the probability $\beta$ is denoted as the transmissibility of the pathogen, and it is defined as the probability per unit time that an $S$ site acquires the infection from a neighbouring $I$ site. Therefore, for a healthy tree with $n$ infected neighbours, the probability of remaining unaffected at each time step is given by $1-(1-\beta)^n$. For simplicity, the neighbourhood is defined by the first four nearest neighbours in the lattice, i.e., a von Neumann neighbourhood.

We consider the limit in which the disease spreads in a much smaller time-scale than the growth of the susceptible species. Moreover, after a patch of forest has died, there exists the possibility of invasion from another species of plant, a phenomenon which has been observed in grass-woodland transitions \citep{AbadesMarquet}. As a consequence, it is unlikely that the woodland site regains susceptibility, and thus, we neglect any regenerative process in the simulations.

The parameters $\beta$ and $T$ regulate the evolution of the disease and both are relevant in the model; sampling from a suitable distribution for each parameter would allow to model levels of disease tolerance to the pathogen for each tree, since it has been identified that some plants exhibit little damage despite a high level presence of the pathogen \citep{MPP:MPP12073}.

For simplicity, in our simulations, we consider uniform values for $\beta$ and $T$. This implies that we are free to set the time-scale by fixing a value for either variable. The time evolution of the landscape is carried out in discrete unitary time intervals. Therefore, by setting $T = 10$, a unitary time interval corresponds to $0.1 T$. This leaves the average vegetation density, $\rho$, and the transmissibility of the pathogen, $\beta$, as the two free parameters.

\section{Results}\label{sec:results}

At the beginning of the simulation, the disease is introduced as a clump of infected trees at the centre of the domain of size $ 5 \times 5$ grid cells, with all trees infected inside this small area. This was found to be sufficient in order to avoid extinction of the disease at initial stages. Under these conditions the transient time, i.e., the time lapse that contains remnants of the initial conditions, was found to be $\sim 200$ time steps. Since we are interested in the steady regime, we discarded this transient from our final calculations. The simulations run until the infected sites disappear or, in order to avoid boundary effects, when infected sites reach any of the four sides. We carried out simulations over $10^4$ ensemble realisations, {\it i.e.} repetitions with different initial conditions with the same tree density $\rho$.

As a first step, we quantify the effect of $\rho$ and $\beta$ in the simulations. At low  $\rho$, the infection quickly dies out, since the distribution of hosts is sparse. On the other side, for higher densities, the epidemic spreads out, infecting most of the trees. Nonetheless, for certain densities, the system shows a critical transition between a self-limited outbreak and a large-scale epiphytotic outbreak. 

\begin{figure}
  \centering
  	\includegraphics[width=0.45\textwidth]{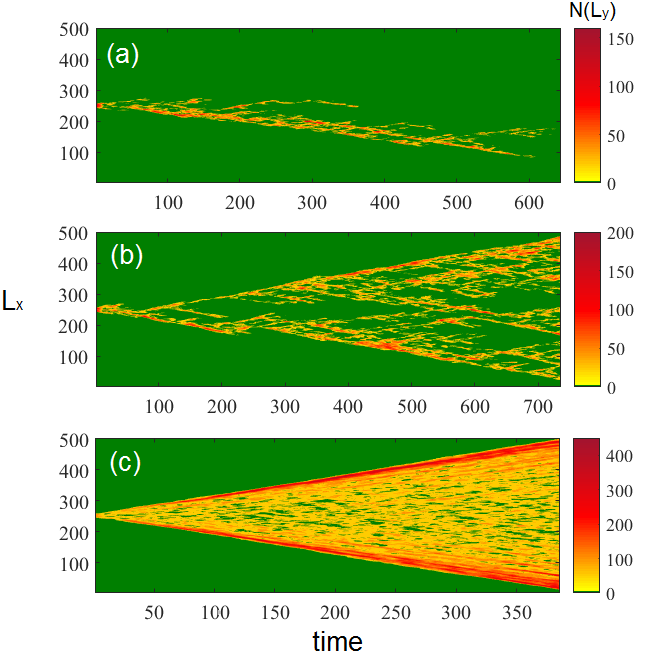}
    \caption{Spatio-temporal behaviour of the number of infected trees $N(L_y)$ along the height of the channel as a function of their location along the width of the domain for the following tree densities $\rho$: (a) 0.58, (b) 0.60 and (c) 0.80. The colour bar shows the number of infected trees per line along $L_y$. The first two cases (a-b) correspond to values in the critical region $\rho \approx \rho_c$ and although the disease is spreading through the domain is not annihilating all susceptible hosts, since there are green sites interspersed with diseased trees. For $\rho=0.8$ the number of infected individuals increases.} 
    \label{spatiotemporal}
\end{figure}

In Fig.~\ref{L500}(a-b), we show the final configuration ($t \sim 600$ time-steps) of the landscape for densities below and above the critical transition. 
Near the critical transition, the pathogen spreads through the domain generating branching structures, and patches of surviving trees may remain unaffected. To highlight this result, the Fig.~\ref{spatiotemporal} shows the spatio-temporal behaviour of the total number of infected hosts, at every time step, found along the direction $N(L_y)$ as a function of their position along the $L_x$ direction, following the same parameters as in Fig.~\ref{L500}. For the first case, $\rho = 0.58$ the disease dies out after approximately 600 time steps, but the fractal-like behaviour of disease spread can be observed as ramifications of infected trees even at a density $\rho = 0.6$. For higher densities $\rho =0.8$, this filamentary-like behaviour is lost and we observe a filled pattern of infected trees. We find that a transition in the severity of the disease occurs in the density interval $[0.56,\, 0.64]$.

Near the critical transition, see figures ~\ref{L500}(b) and ~\ref{spatiotemporal}(a-b), the disease does not annihilate all the trees, but rather spreads through the lattice as active clusters of diseased trees, interspersed with healthy individuals.

\subsection{The phase diagram}

The spreading of the disease has the effect of separating two domains, healthy susceptible trees, $S$, and dead trees, $R$ by a transient interface of infected individuals, $I$. In this model, the number of affected sites is on average, proportional to the landscape area where the infection has been present. By construction, this constant of proportionality is $1/\rho$, and thus, the proportion of affected woodland is

\begin{equation}
	A = \frac{N}{\rho},
	\label{eq:area}
\end{equation}

\noindent
where $N$ is the sum of the $I$ and $R$ sites. 

To quantify the observed dynamics in this system we calculated the spread dynamics of the disease through the lattice via the effective\footnote{We use the term “effective” to emphasize that strictly the velocity cannot be defined in this way close to the critical density since the spanning cluster of diseased trees becomes fractal.  It can be rigorously shown that the velocity with which the disease front propagates is given in 2D as $v \sim \xi^{1-df/d\ell} \sim (p-p_c)^{0.16}$, with $\xi$ denoting the correlation length, $df$ and $d\ell$ the fractal and graph dimensions respectively, \citep{bunde-havlin}. However, Eq. \ref{velocityeq} provides a convenient measure of the rate of spread of the infection, through the epidemic extent or area, which is usually monitored through observational data, \citep{CowgerMundt,mundt2010}.} velocity $v$ of the pathogen. On this basis of Eq.~\ref{eq:area}, a characteristic length-scale, $\mathcal{R} = N^{1/2}$, measures of the radial extent of the disease. Thus, the rate of propagation of the disease in the domain is measured through the velocity $v$, defined as the change in $\mathcal{R}$, i.e.,

\begin{equation}
	v(t) = N^{1/2}(t) - N^{1/2}(t-1).
	\label{velocityeq}
\end{equation}

\begin{figure}
  \centering
    \includegraphics[width=0.45\textwidth]{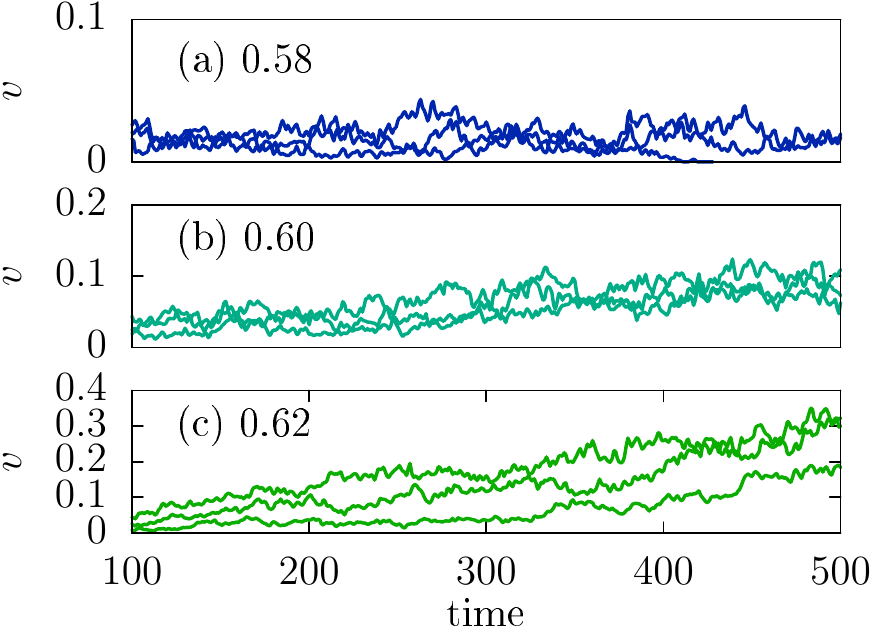}
        \includegraphics[width=0.47\textwidth]{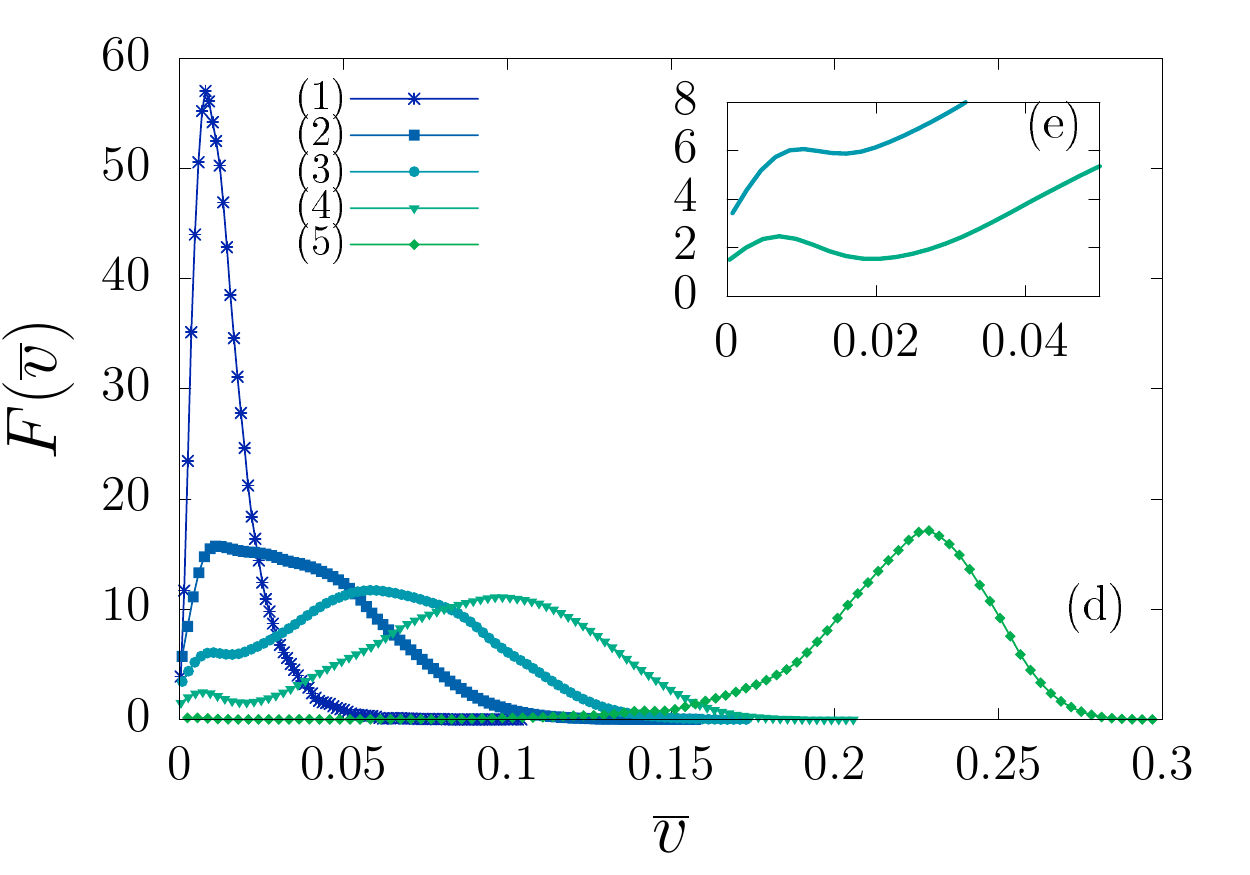}
    \caption{(a-c) Time series for the propagation velocity $v(t)$ for the following values of tree densities: (a) $\rho=0.58$, (b) $\rho=0.60$ and (c) $\rho=0.62$. Three samples are showed for each density. The velocity increases as $\rho$ is increased. (d) Probability distribution function $F(\overline{v})$ for the time averaged velocity $\overline{v}$ obtained from $10^4$ simulations, for densities $\rho$: (1) 0.58, (2) 0.59, (3) 0.595, (4) 0.6 and (5) 0.62. (e) Inset showing the  For all cases $L= 500$ and $\beta = 0.5$. }
    \label{timeseriesvel}
\end{figure}

\noindent
Therefore, Eq.~\ref{velocityeq} measures the spreading velocity in terms of the area of infected trees in the domain. The time series of $v$ is shown in Fig.~\ref{timeseriesvel}(a-c); we will show that the stochasticity observed in these time series gives valuable information about the underlying dynamics when analysed in the framework of early warning indicators for critical transitions.


From the time series for the velocity, we obtain the time average of the velocity $\overline{v}$. Figure \ref{timeseriesvel}(d) shows several probability distribution functions, $F(\overline{v})$, obtained from all realisations, for densities $\rho = 0.58$ (1), 0.59 (2), 0.595 (3), 0.6 (4) and 0.62 (5). For $\rho < \rho_c$, the distribution shows a maximum for $\overline{v} \sim 0.01$, see curve for $\rho = 0.58$. As the tree density increases, and approaches the critical value, $\rho = 0.595$ and $\rho = 0.6$, $F(\overline{v})$ shows clearly that the system can be found in either two states, one for $\overline{v} \approx 0$, which corresponds to local disease confinement and another for $\overline{v} \neq 0$, or epiphytotic outbreak. Figure \ref{timeseriesvel}(e) shows a zoom-in around the local maxima for $\rho$ = 0.595 and 0.6. As the density increases, e.g., $\rho=0.62$, the probability distribution function shows a single maximum for $\overline{v} \approx 0.23$.

After taking the ensemble averages we obtain the mean propagation velocity $\langle \overline{v} \rangle$ as a function of the tree density $\rho$ and various values of the transmission probability $\beta$. We identify from these results a critical density that separates the non-spreading to spreading phase of the disease. The existence of a critical density at $\rho_c$ implies the existence of a spatially connected or spanning cluster of trees for disease spread. From these results, we conclude that this critical density $\rho_c$, is similar in nature to the critical percolation threshold observed in percolation theory \citep{stauffer, gandolfi2013,saberi}, since our computational model only involves a slight modification of the former. 

\begin{figure*}
  \centering
    \includegraphics[width=0.95\textwidth]{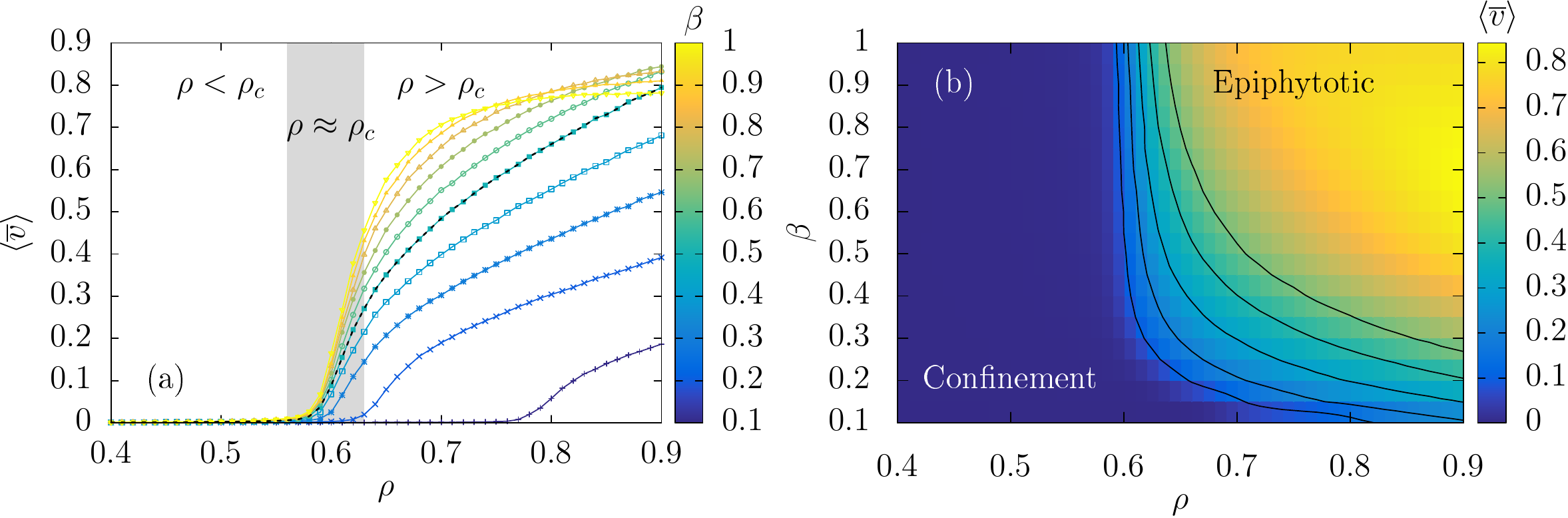}
    \caption{(a) Propagation velocity $\langle \overline{v} \rangle$ for the spread of a pathogen inside a grid with dimensions $L=500$ as a function of tree density $\rho$ and disease transmissibility $\beta$, following a Von Neumann neighbourhood. A shift between two stable states: infection confinement and an extended epiphytotic outbreak, occurs for $\rho \approx \rho_c$. A shaded region is showed for $\rho \sim \rho_c$, associated with the black dotted line highlighting the results for $\beta = 0.5$. (b) Phase space diagram for the pathogen dispersal on the grid, indicating region of disease containment and epiphytotics.} 
    \label{velocity}
\end{figure*}

In Fig.~\ref{velocity}, we observe that the sole effect of the transmission probability $\beta$ = [$0.1$, $1$] is a displacement of the critical point towards lower values of $\rho$. For low disease transmissibility, the density of trees has to be high to have a spanning cluster through the domain. As $\beta$ increases, there is a chance of infecting more trees per infectious period (T) and consequently this cluster occurs at lower densities. Therefore, T, acting conjointly with $\beta$ define the limiting value $\rho_c$. As $\beta$ is increased, the critical transition should tend to the percolation threshold reported in the literature, $\rho_c \rightarrow 0.592746$ \citep{stauffer}. However, since we are working on a finite-size domain, we expect that the critical transition is broadened relative to the result above for infinite-sized domains; in Fig.~\ref{velocity}(a) we highlight this as a shaded region that divides density values according to a region where the critical shift occurs in our simulations (the black dotted line highlights the result for $\beta = 0.5$).

\subsection{Catastrophic shifts in forest disease}

A fundamental emergent property observed in systems near criticality is their capacity to extend over scales comparable to the size of the whole system at long times. Near the critical threshold, short-range interactions lead to the emergence of long-range correlations and the behaviour of the system changes abruptly between two alternative stable states, in this case, local containment and epiphytotics. The occurrence of this shift depends only on the local structure, in our case the density of susceptible hosts. Near the critical transition this system exhibits scale invariance, self-similarity and fractal properties. From the non-stationarity of the time series showed in Fig.~\ref{timeseriesvel}(a-c) we can analyse the underlying dynamics through metric-based indicators proposed in the literature for the identification of early-warning signals: variance, skewness, kurtosis and autocorrelation function at lag 1 \citep{Brock2006, dakos2012, moralesfrank}.

Our goal is to predict the occurrence of a transition between disease containment and epiphytotics using the theory of catastrophic shifts, which in principle could be useful for the prediction of densities at which disease will spread in forests. 

\begin{figure*}
  \centering
    \includegraphics[width=0.95\textwidth]{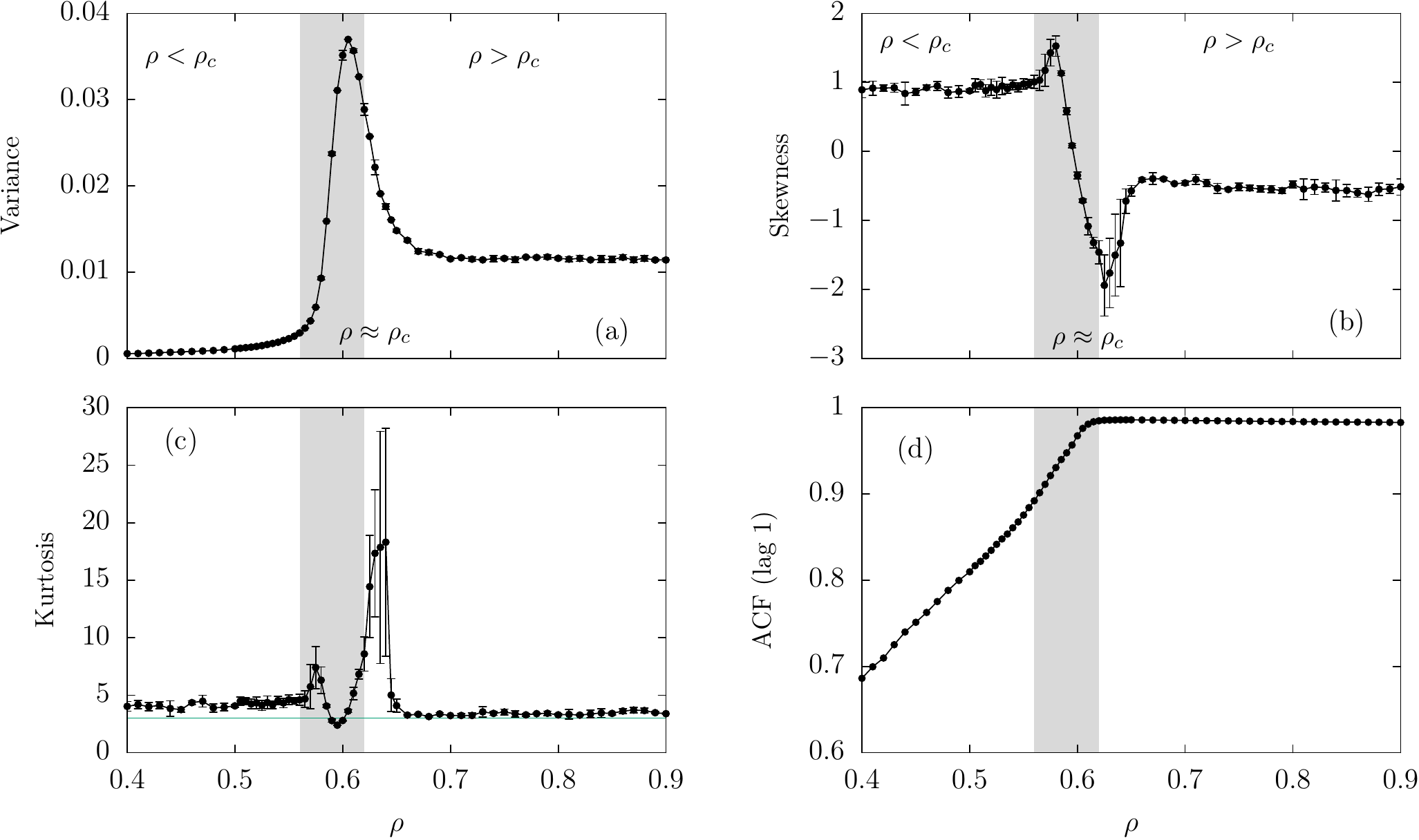}
    \caption{Ensemble behaviour for the metric-based indicators measured for the propagation velocity $v$ of disease spread in the domain ($L=500, \ \beta = 0.5$). Temporal variance (a), kurtosis (b), skewness (c) and autocorrelation function at lag $\tau=1$, as a function of the tree density $\rho$. Three regimes are shown, with the shaded area corresponding to $\rho \approx \rho_c$.}
    \label{stats}
\end{figure*}

We quantify the stochastic variability of $v(t)$ from time series obtained for an ensemble of systems evolving for fixed $\beta = 0.5$ on a domain of size $L = 500$. Our interest was to study the variability in the spreading velocity as the density of trees crosses the critical region. From the probability distribution functions for $\langle \overline{v} \rangle$, we obtained the ensemble behaviour of the following statistical measures: variance (a), kurtosis (b), skewness (c) and autocorrelation function at lag 1 (d), see Fig.~\ref{stats}.

The variance, in Fig.~\ref{stats}(a), shows a rise around the critical point, the increase of this quantity is maximal, and its behaviour is different before the transition occurs, for $\rho < \rho_c$ and after it has happened, $\rho > \rho_c$. The square-root of the variance, the standard deviation, is maximal at the critical transition, which for this finite-size system is $\rho^\text{max} = 0.61$. Therefore, this quantity is useful as an indicator for the prediction of a shift between the disease confinement and epidemics.

The skewness, defined as the third moment of the distribution, quantifies the asymmetry of fluctuations in the time series. It is a useful measure for the prediction of the catastrophic shift since its value changes before and after the transition, depending on whether the system settles down to an alternative state in which the disease propagation is larger or smaller than in the current state, \citep{guttal,dakos2012,kefi}. Our results clearly show both an increase and further decrease in the skewness, see Fig.~\ref{stats}(b). For $\rho < \rho_c$ the skewness is positive and rises up as we approach the critical region. For $\rho \sim \rho_c$, it decreases abruptly and changes sign, becoming negative, i. e., the probability distribution is left-skewed. For higher density of trees, we drive the system away from the critical region, the skewness changes again, and becomes less negative until it settles near zero ($\approx -0.5$). Notably, the rise in skewness observed at $\rho \sim 0.57$, associated with an increase in the nonlinearities of the time series, predicts the outcome of the tipping point. Moreover, this parameter identifies the tree densities for which the system is found in either disease confinement (Skewness $\approx 1$) and epiphytotics (Skewness $\sim 0$).


Strong perturbations can drive the state of a system to reach extreme values close to a transition. Therefore, the probability distribution function of the propagation velocity may show a rise in the kurtosis before the transition is reached. Figure \ref{stats}(c) shows the plot of this quantity obtained in our simulations. The distribution shows two peaks: a local maximum that corresponds to kurtosis values of 7.4 for $\rho \approx 0.57$, and a global maximum with kurtosis of 18.3 for $\rho \approx 0.64$. This indicates that, as the system approaches and exits the critical region, the distribution becomes more strongly peaked, than the reference normal distribution, which has a kurtosis of 3 (blue continuous line), and thus, is leptokurtic. This is consistent with an increased presence of rare values in the propagation velocity. Interestingly, for values closer to the critical point, i. e., $\rho = 0.59$, the kurtosis is 2.4, which is equivalent to a flattened or platykurtic distribution. We conclude that the kurtosis is a good indicator to detect the outcome of the transition.

The temporal autocorrelation function (ACF) measures the spectral properties and changes in the correlation structure, ``memory'', of the time series \citep{dakos2012}. In a general way, the $\tau^\text{th}$ order ACF is defined accordingly as,

\begin{equation}
ACF(\tau) = \frac{\sum_{t=\tau +1}^n (v_t - \overline{v})(v_{t-\tau} - \overline{v})}{\sum_{t=1}^n (v_t - \overline{v})^2}.
\label{ACF}
\end{equation}

\noindent
Following equation \ref{ACF}, we measured the temporal autocorrelation function at lag 1 ($\tau = 1$) in our simulations. Several dynamical systems have shown a slow recovery from small perturbations as they approach the critical transition, phenomenon termed in the literature as ``critical slowing down''. These systems show an increase in the short-term memory of time series which can be detected through an increase of the autocorrelation function at lag 1.

Figure~\ref{stats}(d) show the values for the temporal autocorrelation function at lag 1 measured for the time series of the velocity as a function of the tree density $\rho$. For $\rho < \rho_c$, this quantity increases linearly as we increase the tree density and reaches a maximum threshold inside the critical region for $\rho \geq 0.6$. This is an indication that the system has become increasingly similar between consecutive observations. Since, for $\rho < \rho_c$ there is a fast increase on the ACF, this is useful for the prediction of the outcome of the critical shift in the system.

\section{Discussion}\label{sec:discussion}

The most important question during risk assessment for a forest disease is how pathogens will spread on the landscape, both to predict the occurrence of an epiphytotic outbreak and to assist in designing interventions to counter the onset and progression of the disease. In a real-life scenario, the dispersal of these diseases is complex, mainly due to the multiple geographical and environmental factors affecting the disease spread.

Lattice-based epidemic models have been used previously in the literature to study temporal and spatial fluctuations on the prevalence of epidemic diseases in terms of the minimum tree density for an epidemic to occur \citep{Rhodes1996,Rhodes1997}. The sessility of trees makes lattice modelling of plant diseases more attainable through computational simulations. Works on disease propagation using this framework coupled with historical, geographical and weather information have been used to predict the spread of pathogens through forests on a large scale \citep{HarwoodPautasso,gilligan2010,PotterTomlinson2011,gilligan2012}. These models certainly capture some of the features of previous epiphytotics, and coupling them to the framework of early-warning indicators for detecting critical transitions could be useful for designing strategies against disease spread. 

The following characteristics need to be fulfilled for an epidemic to occur: (i) a critical number of susceptible hosts, (ii) an aggressive phenotype of the pathogen with a high transmissibility rate and (iii) suitable environmental conditions for the pathogen survival. In this paper, we chose to study the effect of the two first factors using a generic stochastic model of epidemic spread on a lattice with a von Neumann neighbourhood. Our model does not incorporate a sophisticated computational description of the system; however it is useful, as a first approximation, for the application of the framework of early warning signals, used widely on complex systems, to reach a new understanding in plant disease epidemics. This could aid in the identification of an optimal planting tree density for the future design of forests, for example, the re-design of the coniferous forests in Scotland, to diminish the impact of disease spread.

Simulations for different tree densities and pathogen transmission indicate a system that shows two stable states: disease confinement and an extended epiphytotic outbreak. We chose to focus our investigations on densities that may result in the system be found in either state. 
All the indicators measured forecast the occurrence of the critical transition. We observe a rise in the variance, skewness and the autocorrelation function at lag 1 as the system approaches $\rho \sim \rho_c$. The skewness also shows a steep change from a positive to a negative value in this region, consistent with the system traversing the critical region and reaching a new stable state. Similarly, the kurtosis, changes from leptokurtic to platykurtic and leptokurtic again in the critical region and immediately afterwards. Consequently, we conclude that all these measures are applicable to predict a transition to epiphytotics. 

Although our current scenario of applicability is a regular domain, far away from the heterogeneous and complex landscapes found in the forests, we hypothesize that their applicability to plant diseases could be fruitful in predicting the outcome of major disease outbreaks \citep{liang-huang}. In real datasets, one of the first challenges would be to define a set of parameters and coarse-grain the system description to an appropriate scale (spatial resolution of the ecological data) to apply these indicators to predict a range of future states of disease propagation. 

Currently, remote sensing technologies, such as satellites and aerial photography are used widely to obtain forest measurements on changes of vegetation index, droughts, fire damage and extent of disease propagation. This information is periodically updated, which implies the availability of spatial datasets taken at time intervals which could be useful to detect the approach to a tipping point before it is crossed. 

In a recent publication \citep{Rogers2018}, several indicators such as the variance, standard deviation, kurtosis and skewness were measured on vegetation indexes (NDVI) time series to detect threshold changes in which the loss of resilience led to state shifts. Their results suggest that that early warning signals of tree mortality are evident up to 24 years and therefore provide a foundation for their potential application on long-term remote sensing data to effectively monitor vegetation patterns and forecast changes in environmental conditions.
Moreover, a study on the quantification of forest fragmentation through aerial images and numerical simulations using a lattice model have suggested that the present state of the tropical forests is close to a critical point of percolation \citep{Taubert18}. Taken together these two studies indicate that the application of the early warning indicators through a lattice model could serve to model and quantify the fragmentation of forests. 

Particularly, the UK has an advantageous position on GIS forest datasets such as the National Forestry Inventory (NFI) \citep{forestryNFIEng}, Light Detection And Ranging (LiDAR) \citep{lidarsurv} and the National Tree  Map\textsuperscript{\textregistered} (NTM)\citep{iTree}, which give accurate information about the woodland patches, 3D forest structure and location and canopy extent of individual trees over 3 m in height, respectively. Moreover, the currently running SAPPHIRE project \citep{sapphireproj}, a collaboration between Forest Research and Rezatec will provide precision maps of tree species and pinpoint trees that exhibit features of stress and disease. Combining all these together, the applicability of early-warning indicators on a complex adaptive system, such as forests, could prove fruitful for devising their stability and resilience to external conditions (such as disease propagation) before a regime shift occurs.

\section*{Acknowledgements}
We thank Dr Willem Roelofs, Dr Alan Macleod and Dr Sam Grant for interesting discussions, and financial support from Defra and Newcastle University, through a Newcastle
University Strategic Impact Award. S. A. Orozco-Fuentes would like to thank to E. R. Gutierrez and A. P. Riascos for comments on early versions of the manuscript.
 
\section*{Data accessibility}

This paper does not use data.

\section*{Author Contributions}
SOF, NGP, RE, and AWB conceived the ideas and designed methodology; SOF implemented the computational model and led the writing of the manuscript. SOF and GG analysed the data. NGP and AWB acquired the funding. All authors contributed critically to the drafts and gave final approval for publication. The authors declare no conflicts of interest.

\bibliographystyle{model2-names.bst}\biboptions{authoryear}

\begin{thebibliography}{52}
\expandafter\ifx\csname natexlab\endcsname\relax\def\natexlab#1{#1}\fi
\providecommand{\url}[1]{\texttt{#1}}
\providecommand{\href}[2]{#2}
\providecommand{\path}[1]{#1}
\providecommand{\DOIprefix}{doi:}
\providecommand{\ArXivprefix}{arXiv:}
\providecommand{\URLprefix}{URL: }
\providecommand{\Pubmedprefix}{pmid:}
\providecommand{\doi}[1]{\href{http://dx.doi.org/#1}{\path{#1}}}
\providecommand{\Pubmed}[1]{\href{pmid:#1}{\path{#1}}}
\providecommand{\bibinfo}[2]{#2}
\ifx\xfnm\relax \def\xfnm[#1]{\unskip,\space#1}\fi
\bibitem[{Abades et~al.()Abades, Gaxiola and Marquet}]{AbadesMarquet}
\bibinfo{author}{Abades, S.R.}, \bibinfo{author}{Gaxiola, A.},
  \bibinfo{author}{Marquet, P.A.}, .
\newblock \bibinfo{title}{Fire, percolation thresholds and the savanna forest
  transition: a neutral model approach}.
\newblock \bibinfo{journal}{Journal of Ecology} \bibinfo{volume}{102},
  \bibinfo{pages}{1386--1393}.
\newblock \URLprefix
  \url{https://besjournals.onlinelibrary.wiley.com/doi/abs/10.1111/1365-2745.12321},
  \DOIprefix\doi{10.1111/1365-2745.12321}.
\bibitem[{Alfinito et~al.(2016)Alfinito, Beccaria and
  Macorini}]{Alfinitosrep27202}
\bibinfo{author}{Alfinito, E.}, \bibinfo{author}{Beccaria, M.},
  \bibinfo{author}{Macorini, G.}, \bibinfo{year}{2016}.
\newblock \bibinfo{title}{Critical behaviour in a stochastic model of vector
  mediated epidemics}.
\newblock \bibinfo{journal}{Scientific Reports} \bibinfo{volume}{6}.
\newblock \URLprefix \url{https://www.nature.com/articles/srep27202},
  \DOIprefix\doi{https://doi.org/10.1038/srep27202}.
\bibitem[{Bak et~al.(1990)Bak, Chen and Tang}]{PerBak}
\bibinfo{author}{Bak, P.}, \bibinfo{author}{Chen, K.}, \bibinfo{author}{Tang,
  C.}, \bibinfo{year}{1990}.
\newblock \bibinfo{title}{A forest-fire model and some thoughts on turbulence}.
\newblock \bibinfo{journal}{Physics Letters A} \bibinfo{volume}{147},
  \bibinfo{pages}{297 -- 300}.
\newblock \URLprefix
  \url{http://www.sciencedirect.com/science/article/pii/037596019090451S},
  \DOIprefix\doi{https://doi.org/10.1016/0375-9601(90)90451-S}.
\bibitem[{Bate et~al.(2016)Bate, Jones, Kleczkowski, MacLeod, Naylor, Timmis,
  Touza and White}]{bate-white}
\bibinfo{author}{Bate, A.M.}, \bibinfo{author}{Jones, G.},
  \bibinfo{author}{Kleczkowski, A.}, \bibinfo{author}{MacLeod, A.},
  \bibinfo{author}{Naylor, R.}, \bibinfo{author}{Timmis, J.},
  \bibinfo{author}{Touza, J.}, \bibinfo{author}{White, P.C.L.},
  \bibinfo{year}{2016}.
\newblock \bibinfo{title}{Modelling the impact and control of an infectious
  disease in a plant nursery with infected plant material inputs}.
\newblock \bibinfo{journal}{Ecological Modelling} \bibinfo{volume}{334},
  \bibinfo{pages}{27--43}.
\newblock \DOIprefix\doi{https://doi.org/10.1016/j.ecolmodel.2016.04.013}.
\bibitem[{Beer and Enting(1990)}]{BeerEnting}
\bibinfo{author}{Beer, T.}, \bibinfo{author}{Enting, I.}, \bibinfo{year}{1990}.
\newblock \bibinfo{title}{Fire spread and percolation modelling}.
\newblock \bibinfo{journal}{Mathematical and Computer Modelling}
  \bibinfo{volume}{13}, \bibinfo{pages}{77 -- 96}.
\newblock \URLprefix
  \url{http://www.sciencedirect.com/science/article/pii/089571779090065U},
  \DOIprefix\doi{https://doi.org/10.1016/0895-7177(90)90065-U}.
\bibitem[{{Bluesky International Ltd}(2017)}]{iTree}
\bibinfo{author}{{Bluesky International Ltd}}, \bibinfo{year}{2017}.
\newblock \bibinfo{title}{National {T}ree {M}ap}.
\newblock \URLprefix
  \url{https://www.blueskymapshop.com/products/national-tree-map}.
\bibitem[{Boulton et~al.(2013)Boulton, Good and Lenton}]{Boulton2013}
\bibinfo{author}{Boulton, C.A.}, \bibinfo{author}{Good, P.},
  \bibinfo{author}{Lenton, T.M.}, \bibinfo{year}{2013}.
\newblock \bibinfo{title}{Early warning signals of simulated {A}mazon
  rainforest dieback}.
\newblock \bibinfo{journal}{Theoretical Ecology} \bibinfo{volume}{6},
  \bibinfo{pages}{373--384}.
\newblock \URLprefix \url{https://doi.org/10.1007/s12080-013-0191-7},
  \DOIprefix\doi{10.1007/s12080-013-0191-7}.
\bibitem[{Bunde and Havlin(1996)}]{bunde-havlin}
\bibinfo{author}{Bunde, A.}, \bibinfo{author}{Havlin, S.},
  \bibinfo{year}{1996}.
\newblock \bibinfo{title}{Fractals and Disoredered Systems}.
\newblock \bibinfo{publisher}{Springer}.
\newblock \DOIprefix\doi{10.1007/978-3-642-84868-1}.
\bibitem[{Carpenter and Brock()}]{Brock2006}
\bibinfo{author}{Carpenter, S.R.}, \bibinfo{author}{Brock, W.A.}, .
\newblock \bibinfo{title}{Rising variance: a leading indicator of ecological
  transition}.
\newblock \bibinfo{journal}{Ecology Letters} \bibinfo{volume}{9},
  \bibinfo{pages}{311--318}.
\newblock \URLprefix
  \url{https://onlinelibrary.wiley.com/doi/abs/10.1111/j.1461-0248.2005.00877.x},
  \DOIprefix\doi{10.1111/j.1461-0248.2005.00877.x},
  \href{http://arxiv.org/abs/https://onlinelibrary.wiley.com/doi/pdf/10.1111/j.1461-0248.2005.00877.x}{\tt
  arXiv:https://onlinelibrary.wiley.com/doi/pdf/10.1111/j.1461-0248.2005.00877.x}.
\bibitem[{Clar et~al.(1996)Clar, Drossel and Schwabl}]{ClarSchwabl}
\bibinfo{author}{Clar, S.}, \bibinfo{author}{Drossel, B.},
  \bibinfo{author}{Schwabl, F.}, \bibinfo{year}{1996}.
\newblock \bibinfo{title}{Forest fires and other examples of self-organized
  criticality}.
\newblock \bibinfo{journal}{Journal of Physics: Condensed Matter}
  \bibinfo{volume}{8}, \bibinfo{pages}{6803 -- 6824}.
\bibitem[{Cobb et~al.(2012)Cobb, Filipe, Meentemeyer, Gilligan and
  Rizzo}]{gilligan2012}
\bibinfo{author}{Cobb, R.C.}, \bibinfo{author}{Filipe, J.A.N.},
  \bibinfo{author}{Meentemeyer, R.K.}, \bibinfo{author}{Gilligan, C.A.},
  \bibinfo{author}{Rizzo, D.M.}, \bibinfo{year}{2012}.
\newblock \bibinfo{title}{Ecosystem transformation by emerging infectious
  disease: loss of large tanoak from {C}alifornia forests}.
\newblock \bibinfo{journal}{Journal of Ecology} ,
  \bibinfo{pages}{712--722}\DOIprefix\doi{10.1111/j.1365-2745.2012.01960.x}.
\bibitem[{Corrado et~al.(2014)Corrado, Cherubini and Pennetta}]{Corrado2014}
\bibinfo{author}{Corrado, R.}, \bibinfo{author}{Cherubini, A.M.},
  \bibinfo{author}{Pennetta, C.}, \bibinfo{year}{2014}.
\newblock \bibinfo{title}{Early warning signals of desertification transitions
  in semiarid ecosystems}.
\newblock \bibinfo{journal}{Phys. Rev. E} \bibinfo{volume}{90},
  \bibinfo{pages}{062705}.
\newblock \URLprefix \url{https://link.aps.org/doi/10.1103/PhysRevE.90.062705},
  \DOIprefix\doi{10.1103/PhysRevE.90.062705}.
\bibitem[{Cowger et~al.(2005)Cowger, Wallace and Mundt}]{CowgerMundt}
\bibinfo{author}{Cowger, C.}, \bibinfo{author}{Wallace, L.D.},
  \bibinfo{author}{Mundt, C.C.}, \bibinfo{year}{2005}.
\newblock \bibinfo{title}{Velocity of spread of wheat stripe rust epidemics}.
\newblock \bibinfo{journal}{Ecology and Epidemiology} \bibinfo{volume}{95},
  \bibinfo{pages}{972--982}.
\newblock \URLprefix
  \url{https://apsjournals.apsnet.org/doi/10.1094/PHYTO-95-0972},
  \DOIprefix\doi{https://doi.org/10.1094/PHYTO-95-0972}.
\bibitem[{Dakos et~al.(2012)Dakos, Carpenter, Brock, Ellison, Guttal, Ives,
  Kéfi, Livina, Seekell, van Nes and Scheffer}]{dakos2012}
\bibinfo{author}{Dakos, V.}, \bibinfo{author}{Carpenter, S.R.},
  \bibinfo{author}{Brock, W.A.}, \bibinfo{author}{Ellison, A.M.},
  \bibinfo{author}{Guttal, V.}, \bibinfo{author}{Ives, A.R.},
  \bibinfo{author}{Kéfi, S.}, \bibinfo{author}{Livina, V.},
  \bibinfo{author}{Seekell, D.A.}, \bibinfo{author}{van Nes, E.H.},
  \bibinfo{author}{Scheffer, M.}, \bibinfo{year}{2012}.
\newblock \bibinfo{title}{Methods for detecting early warnings of critical
  transitions in time series illustrated using simulated ecological data}.
\newblock \bibinfo{journal}{PLOS ONE} \bibinfo{volume}{7},
  \bibinfo{pages}{1--20}.
\newblock \URLprefix \url{https://doi.org/10.1371/journal.pone.0041010},
  \DOIprefix\doi{10.1371/journal.pone.0041010}.
\bibitem[{Drake and Griffen(2010)}]{Griffen2010}
\bibinfo{author}{Drake, J.M.}, \bibinfo{author}{Griffen, B.D.},
  \bibinfo{year}{2010}.
\newblock \bibinfo{title}{Early warning signals of extinction in deteriorating
  environments}.
\newblock \bibinfo{journal}{Nature} \bibinfo{volume}{467},
  \bibinfo{pages}{456--459}.
\newblock \URLprefix \url{https://www.nature.com/articles/nature09389},
  \DOIprefix\doi{https://doi.org/10.1038/nature09389}.
\bibitem[{{Forest Research}(2004)}]{lidarsurv}
\bibinfo{author}{{Forest Research}}, \bibinfo{year}{2004}.
\newblock \bibinfo{title}{Light {D}etection and {R}anging (lidar)}.
\newblock \URLprefix \url{https://www.forestry.gov.uk/fr/lidar}.
\bibitem[{{Forest Research}(2016)}]{forestryNFIEng}
\bibinfo{author}{{Forest Research}}, \bibinfo{year}{2016}.
\newblock \bibinfo{title}{National {F}orestry {I}nventory}.
\newblock \URLprefix \url{https://www.forestry.gov.uk/inventory}.
\bibitem[{{Forest Research}(2018)}]{sapphireproj}
\bibinfo{author}{{Forest Research}}, \bibinfo{year}{2018}.
\newblock \bibinfo{title}{Space {A}pplications for {P}recision {P}lant {H}ealth
  {I}nformation, {R}esponse and {E}valuation (sapphire)}.
\newblock \URLprefix \url{https://www.forestry.gov.uk/fr/sapphire}.
\bibitem[{{Forestry Commission}(2017)}]{forestryShaped}
\bibinfo{author}{{Forestry Commission}}, \bibinfo{year}{2017}.
\newblock \bibinfo{title}{What shaped our forests?}
\newblock \URLprefix \url{https://www.forestry.gov.uk/forestry/infd-5rjhs5}.
\bibitem[{{Forestry Commission}(2018a)}]{forestryPRamorum}
\bibinfo{author}{{Forestry Commission}}, \bibinfo{year}{2018}a.
\newblock \bibinfo{title}{Phytophthora ramorum}.
\newblock \URLprefix \url{https://www.forestry.gov.uk/pramorum}.
\bibitem[{{Forestry Commission}(2018b)}]{forestryNFIScotland}
\bibinfo{author}{{Forestry Commission}}, \bibinfo{year}{2018}b.
\newblock \bibinfo{title}{Phytophthora ramorum}.
\newblock \URLprefix \url{https://www.forestry.gov.uk/datadownload}.
\bibitem[{Gandolfi(2013)}]{gandolfi2013}
\bibinfo{author}{Gandolfi, A.}, \bibinfo{year}{2013}.
\newblock \bibinfo{title}{Percolation Methods for SEIR Epidemics on Graphs}.
  \bibinfo{publisher}{Springer}.
\newblock pp. \bibinfo{pages}{31--58}.
\newblock \URLprefix
  \url{https://app.dimensions.ai/details/publication/pub.1047486272},
  \DOIprefix\doi{10.1007/978-1-4614-9224-5_2}. \bibinfo{note}{exported from
  https://app.dimensions.ai on 2018/11/06}.
\bibitem[{Grassberger(1993)}]{Grassberger93}
\bibinfo{author}{Grassberger, P.}, \bibinfo{year}{1993}.
\newblock \bibinfo{title}{On a self-organized critical forest-fire model}.
\newblock \bibinfo{journal}{Journal of Physics A: Mathematical and General}
  \bibinfo{volume}{26}, \bibinfo{pages}{2081}.
\newblock \URLprefix
  \url{http://iopscience.iop.org/article/10.1088/0305-4470/26/9/007}.
\bibitem[{Gross et~al.(2013)Gross, Holdenrieder, Pautasso, Queloz and
  Sieber}]{MPP:MPP12073}
\bibinfo{author}{Gross, A.}, \bibinfo{author}{Holdenrieder, O.},
  \bibinfo{author}{Pautasso, M.}, \bibinfo{author}{Queloz, V.},
  \bibinfo{author}{Sieber, T.N.}, \bibinfo{year}{2013}.
\newblock \bibinfo{title}{Hymenoscyphus pseudoalbidus, the causal agent of
  european ash dieback}.
\newblock \bibinfo{journal}{Molecular Plant Pathology} \bibinfo{volume}{15},
  \bibinfo{pages}{5--21}.
\newblock \URLprefix
  \url{https://onlinelibrary.wiley.com/doi/abs/10.1111/mpp.12073},
  \DOIprefix\doi{10.1111/mpp.12073},
  \href{http://arxiv.org/abs/https://onlinelibrary.wiley.com/doi/pdf/10.1111/mpp.12073}{\tt
  arXiv:https://onlinelibrary.wiley.com/doi/pdf/10.1111/mpp.12073}.
\bibitem[{Gsell et~al.(2016)Gsell, Scharfenberger, {\"O}zkundakci, Walters,
  Hansson, Janssen, N{\~o}ges, Reid, Schindler, Van~Donk, Dakos and
  Adrian}]{Gsell13122016}
\bibinfo{author}{Gsell, A.S.}, \bibinfo{author}{Scharfenberger, U.},
  \bibinfo{author}{{\"O}zkundakci, D.}, \bibinfo{author}{Walters, A.},
  \bibinfo{author}{Hansson, L.A.}, \bibinfo{author}{Janssen, A.B.G.},
  \bibinfo{author}{N{\~o}ges, P.}, \bibinfo{author}{Reid, P.C.},
  \bibinfo{author}{Schindler, D.E.}, \bibinfo{author}{Van~Donk, E.},
  \bibinfo{author}{Dakos, V.}, \bibinfo{author}{Adrian, R.},
  \bibinfo{year}{2016}.
\newblock \bibinfo{title}{Evaluating early-warning indicators of critical
  transitions in natural aquatic ecosystems}.
\newblock \bibinfo{journal}{Proceedings of the National Academy of Sciences}
  \bibinfo{volume}{113}, \bibinfo{pages}{E8089--E8095}.
\newblock \URLprefix \url{http://www.pnas.org/content/113/50/E8089},
  \DOIprefix\doi{10.1073/pnas.1608242113},
  \href{http://arxiv.org/abs/http://www.pnas.org/content/113/50/E8089.full.pdf}{\tt
  arXiv:http://www.pnas.org/content/113/50/E8089.full.pdf}.
\bibitem[{Guttal and Jayaprakash()}]{guttal}
\bibinfo{author}{Guttal, V.}, \bibinfo{author}{Jayaprakash, C.}, .
\newblock \bibinfo{title}{Changing skewness: an early warning signal of regime
  shifts in ecosystems}.
\newblock \bibinfo{journal}{Ecology Letters} \bibinfo{volume}{11},
  \bibinfo{pages}{450--460}.
\newblock \URLprefix
  \url{https://onlinelibrary.wiley.com/doi/abs/10.1111/j.1461-0248.2008.01160.x},
  \DOIprefix\doi{10.1111/j.1461-0248.2008.01160.x},
  \href{http://arxiv.org/abs/https://onlinelibrary.wiley.com/doi/pdf/10.1111/j.1461-0248.2008.01160.x}{\tt
  arXiv:https://onlinelibrary.wiley.com/doi/pdf/10.1111/j.1461-0248.2008.01160.x}.
\bibitem[{Harwood et~al.(2010)Harwood, Tomlinson, Potter and
  Knight}]{HarwoodPotter2010}
\bibinfo{author}{Harwood, T.D.}, \bibinfo{author}{Tomlinson, I.},
  \bibinfo{author}{Potter, C.A.}, \bibinfo{author}{Knight, J.D.},
  \bibinfo{year}{2010}.
\newblock \bibinfo{title}{Dutch elm disease revisited: past, present and future
  management in great britain}.
\newblock \bibinfo{journal}{Plant Pathology} \bibinfo{volume}{60},
  \bibinfo{pages}{545--555}.
\bibitem[{Kaplan et~al.(2009)Kaplan, Krumhardt and
  Zimmermann}]{forestIndustrial}
\bibinfo{author}{Kaplan, J.O.}, \bibinfo{author}{Krumhardt, K.M.},
  \bibinfo{author}{Zimmermann, N.}, \bibinfo{year}{2009}.
\newblock \bibinfo{title}{The prehistoric and preindustrial deforestation of
  europe}.
\newblock \bibinfo{journal}{Quaternary Science Reviews} \bibinfo{volume}{28},
  \bibinfo{pages}{3016 -- 3034}.
\newblock \URLprefix
  \url{http://www.sciencedirect.com/science/article/pii/S027737910900331X},
  \DOIprefix\doi{https://doi.org/10.1016/j.quascirev.2009.09.028}.
\bibitem[{Kéfi et~al.(2014)Kéfi, Guttal, Brock, Carpenter, Ellison, Livina,
  Seekell, Scheffer, van Nes and Dakos}]{kefi}
\bibinfo{author}{Kéfi, S.}, \bibinfo{author}{Guttal, V.},
  \bibinfo{author}{Brock, W.A.}, \bibinfo{author}{Carpenter, S.R.},
  \bibinfo{author}{Ellison, A.M.}, \bibinfo{author}{Livina, V.N.},
  \bibinfo{author}{Seekell, D.A.}, \bibinfo{author}{Scheffer, M.},
  \bibinfo{author}{van Nes, E.H.}, \bibinfo{author}{Dakos, V.},
  \bibinfo{year}{2014}.
\newblock \bibinfo{title}{Early warning signals of ecological transitions:
  Methods for spatial patterns}.
\newblock \bibinfo{journal}{PLOS ONE} \bibinfo{volume}{9},
  \bibinfo{pages}{1--13}.
\newblock \URLprefix
  \url{https://journals.plos.org/plosone/article?id=10.1371/journal.pone.0092097},
  \DOIprefix\doi{10.1371/journal.pone.0092097}.
\bibitem[{Liang et~al.(2017)Liang, Li, Huang, Qin and Huang}]{liang-huang}
\bibinfo{author}{Liang, L.}, \bibinfo{author}{Li, X.}, \bibinfo{author}{Huang,
  Y.}, \bibinfo{author}{Qin, Y.}, \bibinfo{author}{Huang, H.},
  \bibinfo{year}{2017}.
\newblock \bibinfo{title}{Integrating remote sensing, gis and dynamic models
  for landscape-level simulation of forest insect disturbance}.
\newblock \bibinfo{journal}{Ecological Modelling} \bibinfo{volume}{354},
  \bibinfo{pages}{1--10}.
\newblock \DOIprefix\doi{https://doi.org/10.1016/j.ecolmodel.2017.03.007}.
\bibitem[{Litzow and Hunsicker(2016)}]{Hunsicker2016}
\bibinfo{author}{Litzow, M.A.}, \bibinfo{author}{Hunsicker, M.E.},
  \bibinfo{year}{2016}.
\newblock \bibinfo{title}{Early warning signals, nonlinearity, and signs of
  hysteresis in real ecosystems}.
\newblock \bibinfo{journal}{Ecosphere} \bibinfo{volume}{7}.
\newblock \URLprefix
  \url{https://esajournals.onlinelibrary.wiley.com/doi/full/10.1002/ecs2.1614},
  \DOIprefix\doi{10.1002/ecs2.1614}.
\bibitem[{Macnadbay et~al.(2004)Macnadbay, Bezerra, Fulco, Lyra and
  Argolo}]{MACNADBAY2004249}
\bibinfo{author}{Macnadbay, E.}, \bibinfo{author}{Bezerra, R.},
  \bibinfo{author}{Fulco, U.}, \bibinfo{author}{Lyra, M.},
  \bibinfo{author}{Argolo, C.}, \bibinfo{year}{2004}.
\newblock \bibinfo{title}{Critical behavior of a vector-mediated propagation of
  an epidemic process}.
\newblock \bibinfo{journal}{Physica A: Statistical Mechanics and its
  Applications} \bibinfo{volume}{342}, \bibinfo{pages}{249 -- 255}.
\newblock \URLprefix
  \url{http://www.sciencedirect.com/science/article/pii/S0378437104005059},
  \DOIprefix\doi{https://doi.org/10.1016/j.physa.2004.04.085}.
\bibitem[{Macpherson et~al.(2017)Macpherson, Kleczkowski, Healey, Quine and
  Hanley}]{macpherson-hanley}
\bibinfo{author}{Macpherson, M.F.}, \bibinfo{author}{Kleczkowski, A.},
  \bibinfo{author}{Healey, J.R.}, \bibinfo{author}{Quine, C.P.},
  \bibinfo{author}{Hanley, N.}, \bibinfo{year}{2017}.
\newblock \bibinfo{title}{The effects of invasive pests and pathogens on
  strategies for forest diversification}.
\newblock \bibinfo{journal}{Ecological Modelling} \bibinfo{volume}{350},
  \bibinfo{pages}{87--99}.
\newblock \DOIprefix\doi{https://doi.org/10.1016/j.ecolmodel.2017.02.003}.
\bibitem[{Maskell et~al.(2013)Maskell, Henrys, Norton, Smart and
  Wood}]{AshTreeDistro}
\bibinfo{author}{Maskell, L.}, \bibinfo{author}{Henrys, P.},
  \bibinfo{author}{Norton, L.}, \bibinfo{author}{Smart, S.},
  \bibinfo{author}{Wood, C.}, \bibinfo{year}{2013}.
\newblock \bibinfo{title}{Distribution of ash trees ({F}raxinus excelsior) in
  countryside survey data}.
\newblock \bibinfo{journal}{Country Side Survey} ,
  \bibinfo{pages}{1--20}\URLprefix \url{www.countrysidesurvey.org.uk}.
\bibitem[{Meentemeyer et~al.(2010)Meentemeyer, Cunniffe, Cook, Filipe, Hunter,
  Rizzo and Gilligan}]{gilligan2010}
\bibinfo{author}{Meentemeyer, R.K.}, \bibinfo{author}{Cunniffe, N.J.},
  \bibinfo{author}{Cook, A.R.}, \bibinfo{author}{Filipe, J.A.N.},
  \bibinfo{author}{Hunter, R.D.}, \bibinfo{author}{Rizzo, D.M.},
  \bibinfo{author}{Gilligan, C.A.}, \bibinfo{year}{2010}.
\newblock \bibinfo{title}{Epidemiological modeling of invasion in heterogeneous
  landscapes: spread of sudden oak death in california (1990-2030)}.
\newblock \bibinfo{journal}{Ecosphere} \bibinfo{volume}{2},
  \bibinfo{pages}{1--24}.
\bibitem[{Mitchell et~al.(2014)Mitchell, Beaton, Bellamy, Broome, Chetcuti,
  Eaton, Ellis, Gimona, Harmer, Hester, Hewison, Hodgetts, Iason, Kerr,
  Littlewood, Newey, Potts, Pozsgai, Ray, Sim, Stockan, Taylor and
  Woodward}]{Mitchell201495}
\bibinfo{author}{Mitchell, R.}, \bibinfo{author}{Beaton, J.},
  \bibinfo{author}{Bellamy, P.}, \bibinfo{author}{Broome, A.},
  \bibinfo{author}{Chetcuti, J.}, \bibinfo{author}{Eaton, S.},
  \bibinfo{author}{Ellis, C.}, \bibinfo{author}{Gimona, A.},
  \bibinfo{author}{Harmer, R.}, \bibinfo{author}{Hester, A.},
  \bibinfo{author}{Hewison, R.}, \bibinfo{author}{Hodgetts, N.},
  \bibinfo{author}{Iason, G.}, \bibinfo{author}{Kerr, G.},
  \bibinfo{author}{Littlewood, N.}, \bibinfo{author}{Newey, S.},
  \bibinfo{author}{Potts, J.}, \bibinfo{author}{Pozsgai, G.},
  \bibinfo{author}{Ray, D.}, \bibinfo{author}{Sim, D.},
  \bibinfo{author}{Stockan, J.}, \bibinfo{author}{Taylor, A.},
  \bibinfo{author}{Woodward, S.}, \bibinfo{year}{2014}.
\newblock \bibinfo{title}{Ash dieback in the uk: A review of the ecological and
  conservation implications and potential management options}.
\newblock \bibinfo{journal}{Biological Conservation} \bibinfo{volume}{175},
  \bibinfo{pages}{95 -- 109}.
\newblock \URLprefix
  \url{http://www.sciencedirect.com/science/article/pii/S0006320714001700},
  \DOIprefix\doi{https://doi.org/10.1016/j.biocon.2014.04.019}.
\bibitem[{Morales et~al.(2015)Morales, Landa, Angeles, Toledo, Rivera, Temis
  and Frank}]{moralesfrank}
\bibinfo{author}{Morales, I.O.}, \bibinfo{author}{Landa, E.},
  \bibinfo{author}{Angeles, C.C.}, \bibinfo{author}{Toledo, J.C.},
  \bibinfo{author}{Rivera, A.L.}, \bibinfo{author}{Temis, J.M.},
  \bibinfo{author}{Frank, A.}, \bibinfo{year}{2015}.
\newblock \bibinfo{title}{Behavior of early warnings near the critical
  temperature in the two-dimensional ising model}.
\newblock \bibinfo{journal}{PLOS ONE} \bibinfo{volume}{10},
  \bibinfo{pages}{1--20}.
\newblock \URLprefix \url{https://doi.org/10.1371/journal.pone.0130751},
  \DOIprefix\doi{10.1371/journal.pone.0130751}.
\bibitem[{Mundt et~al.(2013)Mundt, Wallace, Allen, Hollier, Kemerait and
  Sikora}]{mundt2010}
\bibinfo{author}{Mundt, C.C.}, \bibinfo{author}{Wallace, L.D.},
  \bibinfo{author}{Allen, T.W.}, \bibinfo{author}{Hollier, C.A.},
  \bibinfo{author}{Kemerait, R.C.}, \bibinfo{author}{Sikora, E.J.},
  \bibinfo{year}{2013}.
\newblock \bibinfo{title}{Initial epidemic area is strongly associated with the
  yearly extent of soybean rust spread in north america}.
\newblock \bibinfo{journal}{Biol. Invasions} \bibinfo{volume}{15},
  \bibinfo{pages}{1431--1438}.
\bibitem[{Potter et~al.(2011)Potter, Harwood, Knight and
  Tomlinson}]{PotterTomlinson2011}
\bibinfo{author}{Potter, C.}, \bibinfo{author}{Harwood, T.},
  \bibinfo{author}{Knight, J.}, \bibinfo{author}{Tomlinson, I.},
  \bibinfo{year}{2011}.
\newblock \bibinfo{title}{Learning from history, predicting the future: the
  {UK} dutch elm disease outbreak in relation to contemporary tree disease
  threats}.
\newblock \bibinfo{journal}{Philosophical Transactions of the Royal Society B}
  \bibinfo{volume}{366}, \bibinfo{pages}{1966--1974}.
\bibitem[{Ratajczak et~al.()Ratajczak, D'Odorico, Nippert, Collins, Brunsell
  and Ravi}]{Ravi2017}
\bibinfo{author}{Ratajczak, Z.}, \bibinfo{author}{D'Odorico, P.},
  \bibinfo{author}{Nippert, J.B.}, \bibinfo{author}{Collins, S.L.},
  \bibinfo{author}{Brunsell, N.A.}, \bibinfo{author}{Ravi, S.}, .
\newblock \bibinfo{title}{Changes in spatial variance during a grassland to
  shrubland state transition}.
\newblock \bibinfo{journal}{Journal of Ecology} \bibinfo{volume}{105},
  \bibinfo{pages}{750--760}.
\newblock \URLprefix
  \url{https://besjournals.onlinelibrary.wiley.com/doi/abs/10.1111/1365-2745.12696},
  \DOIprefix\doi{10.1111/1365-2745.12696},
  \href{http://arxiv.org/abs/https://besjournals.onlinelibrary.wiley.com/doi/pdf/10.1111/1365-2745.12696}{\tt
  arXiv:https://besjournals.onlinelibrary.wiley.com/doi/pdf/10.1111/1365-2745.12696}.
\bibitem[{Rhodes and Anderson(1996)}]{Rhodes1996}
\bibinfo{author}{Rhodes, C.}, \bibinfo{author}{Anderson, R.},
  \bibinfo{year}{1996}.
\newblock \bibinfo{title}{Dynamics in a lattice epidemic model}.
\newblock \bibinfo{journal}{Physics Letters A} \bibinfo{volume}{210},
  \bibinfo{pages}{183 -- 188}.
\newblock \URLprefix
  \url{http://www.sciencedirect.com/science/article/pii/S0375960196800077},
  \DOIprefix\doi{https://doi.org/10.1016/S0375-9601(96)80007-7}.
\bibitem[{Rhodes and Anderson(1997)}]{Rhodes1997}
\bibinfo{author}{Rhodes, C.}, \bibinfo{author}{Anderson, R.},
  \bibinfo{year}{1997}.
\newblock \bibinfo{title}{Epidemic thresholds and vaccination in a lattice
  model of disease spread}.
\newblock \bibinfo{journal}{Theoretical Population Biology}
  \bibinfo{volume}{52}, \bibinfo{pages}{101 -- 118}.
\newblock \URLprefix
  \url{http://www.sciencedirect.com/science/article/pii/S004058099791323X},
  \DOIprefix\doi{https://doi.org/10.1006/tpbi.1997.1323}.
\bibitem[{Rist and Moen(2013)}]{rist-moen}
\bibinfo{author}{Rist, L.}, \bibinfo{author}{Moen, J.}, \bibinfo{year}{2013}.
\newblock \bibinfo{title}{Sustainability in forest management and a new role
  for resilience thinking}.
\newblock \bibinfo{journal}{Forest Ecology and Management}
  \bibinfo{volume}{310}.
\newblock \DOIprefix\doi{http://dx.doi.org/10.1038/ncomms2632}.
\bibitem[{Rogers et~al.()Rogers, Solvik, Hogg, Ju, Masek, Michaelian, Berner
  and Goetz}]{Rogers2018}
\bibinfo{author}{Rogers, B.M.}, \bibinfo{author}{Solvik, K.},
  \bibinfo{author}{Hogg, E.H.}, \bibinfo{author}{Ju, J.},
  \bibinfo{author}{Masek, J.G.}, \bibinfo{author}{Michaelian, M.},
  \bibinfo{author}{Berner, L.T.}, \bibinfo{author}{Goetz, S.J.}, .
\newblock \bibinfo{title}{Detecting early warning signals of tree mortality in
  boreal north america using multiscale satellite data}.
\newblock \bibinfo{journal}{Global Change Biology} \bibinfo{volume}{24},
  \bibinfo{pages}{2284--2304}.
\newblock \URLprefix
  \url{https://onlinelibrary.wiley.com/doi/abs/10.1111/gcb.14107},
  \DOIprefix\doi{10.1111/gcb.14107},
  \href{http://arxiv.org/abs/https://onlinelibrary.wiley.com/doi/pdf/10.1111/gcb.14107}{\tt
  arXiv:https://onlinelibrary.wiley.com/doi/pdf/10.1111/gcb.14107}.
\bibitem[{Saberi(2015)}]{saberi}
\bibinfo{author}{Saberi, A.A.}, \bibinfo{year}{2015}.
\newblock \bibinfo{title}{Recent advances in percolation theory and its
  applications}.
\newblock \bibinfo{journal}{Physics Reports} \bibinfo{volume}{578},
  \bibinfo{pages}{1 -- 32}.
\newblock \URLprefix
  \url{http://www.sciencedirect.com/science/article/pii/S0370157315002008},
  \DOIprefix\doi{https://doi.org/10.1016/j.physrep.2015.03.003}.
  \bibinfo{note}{recent advances in percolation theory and its applications}.
\bibitem[{Scheffer(2009)}]{scheffer-book}
\bibinfo{author}{Scheffer, M.}, \bibinfo{year}{2009}.
\newblock \bibinfo{title}{Critical Transitions in Nature and Society}.
\newblock \bibinfo{publisher}{Princeton University Press}.
\bibitem[{Scheffer et~al.(2009)Scheffer, Bascompte, Brock, Brovkin, Carpenter,
  Dakos, Held, van Nes, Rietkerk and Sugihara}]{schefferNature}
\bibinfo{author}{Scheffer, M.}, \bibinfo{author}{Bascompte, J.},
  \bibinfo{author}{Brock, W.A.}, \bibinfo{author}{Brovkin, V.},
  \bibinfo{author}{Carpenter, S.R.}, \bibinfo{author}{Dakos, V.},
  \bibinfo{author}{Held, H.}, \bibinfo{author}{van Nes, E.H.},
  \bibinfo{author}{Rietkerk, M.}, \bibinfo{author}{Sugihara, G.},
  \bibinfo{year}{2009}.
\newblock \bibinfo{title}{Early-warning signals for critical transitions}.
\newblock \bibinfo{journal}{Nature} \bibinfo{volume}{461}.
\bibitem[{Schneider et~al.(2012)Schneider, Rasband and Eliceiri}]{Imagej}
\bibinfo{author}{Schneider, C.}, \bibinfo{author}{Rasband, W.},
  \bibinfo{author}{Eliceiri, K.}, \bibinfo{year}{2012}.
\newblock \bibinfo{title}{Nih image to imagej: 25 years of image analysis}.
\newblock \bibinfo{journal}{Nature Methods} ,
  \bibinfo{pages}{671--675}\URLprefix
  \url{https://www.nature.com/articles/nmeth.2089}.
\bibitem[{Stauffer and Aharony(2003)}]{stauffer}
\bibinfo{author}{Stauffer, D.}, \bibinfo{author}{Aharony, A.},
  \bibinfo{year}{2003}.
\newblock \bibinfo{title}{Introduction to Percolation Theory}.
\newblock \bibinfo{edition}{Revised} ed., \bibinfo{publisher}{Taylor \&
  Francis}.
\bibitem[{Taubert et~al.(2018)Taubert, Fischer, Groeneveld, Lehmann, Müller,
  Rödig, Wiegand and Huth}]{Taubert18}
\bibinfo{author}{Taubert, F.}, \bibinfo{author}{Fischer, R.},
  \bibinfo{author}{Groeneveld, J.}, \bibinfo{author}{Lehmann, S.},
  \bibinfo{author}{Müller, M.S.}, \bibinfo{author}{Rödig, E.},
  \bibinfo{author}{Wiegand, T.}, \bibinfo{author}{Huth, A.},
  \bibinfo{year}{2018}.
\newblock \bibinfo{title}{Global patterns of tropical forest fragmentation}.
\newblock \bibinfo{journal}{Nature} , \bibinfo{pages}{519--522}\URLprefix
  \url{https://www.nature.com/articles/nature25508}.
\bibitem[{Thomas(2016)}]{JEC:JEC12566}
\bibinfo{author}{Thomas, P.A.}, \bibinfo{year}{2016}.
\newblock \bibinfo{title}{Biological flora of the {B}ritish {I}sles: Fraxinus
  excelsior}.
\newblock \bibinfo{journal}{Journal of Ecology} \bibinfo{volume}{104},
  \bibinfo{pages}{1158--1209}.
\bibitem[{Xu et~al.()Xu, Harwood, Pautasso and Jeger}]{HarwoodPautasso}
\bibinfo{author}{Xu, X.}, \bibinfo{author}{Harwood, T.D.},
  \bibinfo{author}{Pautasso, M.}, \bibinfo{author}{Jeger, M.J.}, .
\newblock \bibinfo{title}{Spatio-temporal analysis of an invasive plant
  pathogen (phytophthora ramorum) in england and wales}.
\newblock \bibinfo{journal}{Ecography} \bibinfo{volume}{32},
  \bibinfo{pages}{504--516}.
\newblock \URLprefix
  \url{https://onlinelibrary.wiley.com/doi/abs/10.1111/j.1600-0587.2008.05597.x},
  \DOIprefix\doi{10.1111/j.1600-0587.2008.05597.x},
  \href{http://arxiv.org/abs/https://onlinelibrary.wiley.com/doi/pdf/10.1111/j.1600-0587.2008.05597.x}{\tt
  arXiv:https://onlinelibrary.wiley.com/doi/pdf/10.1111/j.1600-0587.2008.05597.x}.

\end{thebibliography}

\end{document}